\documentclass[prd,nofootinbib,floats,superscriptaddress,eqsecnum,tightenlines,preprintnumbers,11pt]{revtex4}

\usepackage{hyperref}
\usepackage{graphicx}
\usepackage{amsmath,amssymb,amsfonts,amsthm,latexsym}
\usepackage{marginnote}
\usepackage{color}

\def\beq{\begin{equation}}
\def\eeq{\end{equation}}
\newcommand{\bea}{\begin{eqnarray}}
\newcommand{\eea}{\end{eqnarray}}
\def\bi{\begin{itemize}}
\def\ei{\end{itemize}}
\def\ba{\begin{array}}
\def\ea{\end{array}}
\def\bfig{\begin{figure}}
\def\efig{\end{figure}}

\begin{document}

\title{Minimally Modified Gravity: a Hamiltonian Construction}
\author{S. Mukohyama}
\affiliation{Center for Gravitational Physics, Yukawa Institute for Theoretical Physics, Kyoto University, 606-8502, Kyoto, Japan}
\affiliation{Kavli Institute for the Physics and Mathematics of the Universe (WPI),
The University of Tokyo Institutes for Advanced Study,\\
The University of Tokyo, Kashiwa, Chiba 277-8583, Japan}
\author{K. Noui}
\affiliation{Institut Denis Poisson, CNRS, Universit\'e de Tours, Universit\'e d'Orl\'eans, France}
\affiliation{Laboratoire Astroparticule et Cosmologie, Universit\'e Paris Diderot, France}

\date{\today}

\begin{abstract}
Minimally modified gravity theories are modifications of general relativity with two local gravitational degrees of freedom in four dimensions.
Their construction relies on the breaking of 4D diffeomorphism invariance keeping however the symmetry under 3D
diffeomorphisms. Here, we construct these theories from a Hamiltonian point of view. We start with the phase space of general relativity
in the ADM formalism. Then, we find the conditions that the Hamiltonian must satisfy for the theory to propagate (up to) two gravitational
degrees of freedom with the assumptions that the lapse and the shift are not dynamical, and the theory remains invariant under 3D diffeomorphisms.
This construction enables us to recover the well-known ``cuscuton" class of scalar-tensor theories in the unitary gauge. 
We also exhibit a new class of interesting theories, that we dubb $f({\cal H})$ theories, where 
 the usual Hamiltonian constraint $\cal H$ of general relativity is replaced by $f({\cal H})$ where $f$ is an arbitrary function.
\end{abstract}

\maketitle

\section{Introduction}
A century after its discovery, the theory of general relativity continues to challenge all validity tests.
The latest is the fabulous detection of gravitational waves emitted by a neutron star merger \cite{TheLIGOScientific:2017qsa} with a first measurement of their  propagation speed which is probably the same as the speed of light in vacuum... as predicted by Einstein. In spite of all these successes, the reasons for believing that general relativity is not the ultimate theory of space-time and that 
it will have to be surpassed are numerous and so interesting that modifying gravity has become a  very dynamical field of research per se in theoretical physics and cosmology
these last years (see \cite{Clifton:2011jh,Koyama:2015vza} for example).

Going beyond general relativity necessary leads to relax one of the fundamental hypothesis of the Lovelock theorem 
that makes Einstein theory unique: invariance under diffeomorphisms, locality, pure
metric formulation in four space-time dimensions... For instance, massive gravity in four dimensions \cite{deRham:2010kj} renounces to the invariance under diffeomorphims, and scalar-tensor theories relies on the fact 
that a scalar degree of freedom comes with the metric to describe the physics of space-time (at  least at very large or very short scales)... There exist many  modifications 
of gravity, and most of them (exactly like massive gravity or scalar-tensor theories) often share the property that one or more additional degree(s) of freedom propagate in the theory. When such theories are designed to account for dark energy for example, the extra degrees of freedom are responsible for the fifth force which makes the expansion of the universe accelerating. When they are constructed 
to cure the well-known (in)famous ultra-violet problems of general relativity, these degrees of freedom play the role of making the theory renormalizable, and eventually  quantifying
\cite{Horava:2009uw,Blas:2010hb}.
Hence, there might be the general belief that modifying gravity cannot be done without introducing new degree(s) of freedom in the scenario, in addition to the usual two spin-2
massless degrees of freedom of the gravitational field. 

In this spirit, scalar-tensor theories are sometimes considered as  the ``simplest" theories of modified gravity because they come with 
one extra degree of freedom only.  These last years, they have been at the core of a huge activity and scalar-tensor theories, whose
 actions involve up to second derivatives of the scalar field, have been systematically classified and extensively studied \cite{Horndeski:1974wa,Deffayet:2011gz,Kobayashi:2011nu,Gleyzes:2014dya,Gleyzes:2014qga,Lin:2014jga,Langlois:2015cwa,Langlois:2015skt,Crisostomi:2016tcp,Crisostomi:2016czh,Deffayet:2015qwa,Achour:2016rkg,BenAchour:2016fzp,Langlois:2017mxy}.   
 Adding higher order derivatives in a Lagrangian is potentially very dangerous because it could lead to the fact that, not only one, but two scalars propage in the theory,
 one of them being the  Ostrogradski ghost. Degeneracy conditions \cite{Langlois:2015cwa} 
 in a higher order scalar-tensor theory insure that at most three degrees of freedom propagate, but one has
 to study the theory in more details to see whether these degrees of freedom are safe or not. 
 Furthermore, it has been realized that degeneracy conditions in the unitary gauge (where the
 scalar field is fixed to be a function of time only) is enough to  ensure that a unique scalar propagates in addition to the usual two tensor modes \cite{DeFelice:2018mkq}.  Also, there exists the possibility that only two tensorial degrees of freedom propagate  in a scalar-tensor theory which is, of course, different from gravity (the scalar mode is in fact shadowy in the sense of \cite{DeFelice:2018mkq}). 
They form the class of ``cuscuton" theories \cite{Afshordi:2006ad,Iyonaga:2018vnu}.
 These theories are particularly interesting and they can be
 considered as minimal modifications of general relativity. 

A systematic construction of gravitational theories with only (up to) two degrees of freedom has been initiated in \cite{Lin:2017oow,Aoki:2018brq}. 
The  idea consists in renouncing to the invariance under four dimensional diffeomorphisms but keeping  the three dimensional diff-invariance. 
This is equivalent to considering scalar-tensor theories in the unitary gauge. As generically 
Lorentz-breaking gravity have more than two degrees of freedom, one has to find the conditions for 
the theory to possess enough constraints that would  kill the extra degrees of freedom, which would leave us with (at most) two gravitational degrees modes. More precisely, one starts with
the ADM parametrization of the metric
\bea
\label{ADM}
ds^2 = - N^2 dt^2 + h_{ij} (dx^i + N^i dt)(dx^j + N^j dt) \, ,
\eea 
where $N$, $N^i$ and $h_{ij}$ are respectively the lapse function, the shift vector and the  induced spatial metric. Then, one considers general
actions of the form
\bea
\label{general action}
S[N,N^i,h_{ij}] = \int d^3x \, dt \, \sqrt{h} \, {\cal L}(K_{ij},R_{ij},h^{ij},N,\nabla_i) \, ,
\eea
 where $K_{ij}$ is the extrinsic curvature, $R_{ij}$ the three-dimensional curvature and $\nabla_i$ the spatial covariant derivative.
 And finally, one performs a Hamiltonian analysis to find the necessary conditions for the theory to propagate (at most) two degrees of freedom.
 This program was completed in the case where the Lagrangian \eqref{general action} was supposed to be linear in the lapse function \cite{Lin:2017oow}.
 In that way, one found a large class of modified theory of gravity with only two degrees of freedom that have been dubbed for obvious reasons ``minimally modified gravity". 

In this paper, we construct minimally modified gravity from a Hamiltonian point of view with the idea the Hamiltonian framework is more suited for studying and
classifying Lorentz breaking theories than the Lagrangian framework. 
Indeed, we modify the phase space of general relativity (and not directly
the Lagrangian) in such a way that the modified theory remains invariant under spatial diffeomorphisms only and still propagates two tensorial degrees of freedom. 
More precisely, we start with a phase space which is parametrized by the usual ten pairs of conjugate variables (the metric variables in the ADM decomposition and their momenta),
and we consider a ``modified" Hamiltonian of the form
\bea
\label{H intro}
H = \int d^3x \, \sqrt{h} \, \left[ {\cal H}(\pi^{ij},R_{ij},h^{ij},N,\nabla_i) + N^i {\cal V}_i \right] \, ,
\eea
where ${\cal V}_i$ is the usual vectorial constraint, and ${\cal H}$ is a three dimensional diff-invariant function which is a priori different from the usual scalar constraint.
Then, the problem consists in finding the conditions that $\cal H$ must satisfy for the theory to propagate  two (or less) degrees of freedom. We address this issue and
find that ${\cal H}$ must be an affine function of the lapse, of the form 
\bea
\label{cond1}
{\cal H}= N \, {\cal H}_0(\pi^{ij},R_{ij},h^{ij},\nabla_i)  + {\cal V}(\pi^{ij},R_{ij},h^{ij},\nabla_i) \, , 
\eea 
with additional conditions on the functions ${\cal H}_0$ and $\cal V$. {A necessary condition is that $\{  {\cal H}_0(x) ,  {\cal H}_0(y)\}$, viewed as an operator
acting on the space of functions Fun($M$) on the space manifold by integration, has a non-trivial kernel, and a sufficient condition is that }
\bea
\label{cond2}
\{  {\cal H}_0(x) ,  {\cal H}_0(y)\} \, \approx \, 0 \, ,
\eea 
where $\approx$ means weakly vanishing (i.e. it vanishes up to constraints). 
In this construction, we recover the well-known class of ``cuscuton'' theories that can be extended to non-local theories. But we also find new classes of theories.
In particular, we exhibit a remarkably simple class of theories which are such that ${\cal H}_0=  f({\cal H}_{gr})$ where
 $f$ is an arbitrary function and ${\cal H}_{gr}$ is the usual scalar constraint of general relativity. 
 Such theories are invariant under a four dimensional local symmetry (which contains the 3D diffeomorphims) and possess very interesting properties that we discuss in the paper. 

\medskip

The paper is organized as follows. We start, in section \ref{Maxwell} with the simpler case of a spin-1 field to illustrate our construction. 
Hence, we construct modified Maxwell theory in a four dimensional Minkowski space-time, where the dynamical variable is a one form $A_\mu$. 
To mimic the construction of minimally modified theories of gravity, we relax some hypothesis which makes Maxwell theory unique:
we break the $U(1)$ gauge symmetry and also the global Lorentz invariance keeping, however, a symmetry under one rotational subgroup $SO(3)$ (the one
that leaves $A_0$ invariant).
Then, we modify the Maxwell Hamiltonian and find the conditions for the new theory to propagate only (up to) two degrees of freedom.
Finally, we give some concrete examples. In section \ref{gravity}, we turn to the more interesting case of minimally modified gravities. 
We write  conditions that the modified Hamiltonian constraint \eqref{H intro} must satisfy to have (up to) two tensorial degrees of freedom. 
These conditions \eqref{cond1} and \eqref{cond2} appear to be very simple in the Hamiltonian framework, and they can be explicitly solved in some cases. As we said previously, we recover the cuscuton theories, and we find an interesting and remarkably simple new class of theories, dubbed $f({\cal H})$ theories, 
where the usual Hamiltonian constraint of general relativity ${\cal H}_{gr}$ has been replaced by $f({\cal H}_{gr})$ where $f$ is an arbitrary function. 
We quickly study their cosmology %and their Newton limit 
to show interesting differences with general relativity.
We conclude with a brief summary of  our results and some perspectives. 

\section{Minimally Modified Maxwell Theory}
\label{Maxwell}
Following the ideas that lead to the construction of minimally modified gravity theories, 
we  build, in this section, a large class of modified Maxwell theories which propagates 2 (vectorial) degrees of freedom in the 4-dimensional Minkowski space-time.
Maxwell theory provides us with a simpler but very interesting context to illustrate the construction of minimally modified gravity theories from a Hamiltonian point of view that
we will present in section \ref{gravity}.

\subsection{Framework: symmetry breaking and degeneracy}
Maxwell theory
is the unique free action for a $U(1)$ connection $A_\mu$ evolving in a Minkowski space-time, which is invariant under the usual $U(1)$ gauge symmetry, also invariant under the global Lorentz symmetry (i.e.  the isometry group of the Minkowski metric $SO(1,3)$), and which in addition produces (at most) second order equations of motion. The $U(1)$ invariance implies that the action is a functional  of the curvature two-form only
\bea
\label{curvature}
F_{\mu\nu} \; \equiv \; \partial_\mu A_\nu - \partial_\nu A_\mu \, .
\eea
The global Lorentz symmetry implies that the curvature components must be contracted with the metric $\eta_{\mu\nu}= \text{diag}(-1,1,1,1)$ (and its inverse) 
such that the Lagrangian density is a scalar for the Lorentz group. Finally, the freeness of the theory says that
the action is at most quadratic in the connection. Hence  the only possible theory is described by the action (in vacuum)
\bea
S[A_\mu] \; = \; -\frac{1}{4 \mu_0} \int d^4x \, F_{\mu\nu} F^{\mu\nu},
\eea
where $\mu_0$ is the usual permeability, and indices are raised with $\eta^{\mu\nu}$. 
A simple analysis shows that this very well-known theory propagates only 2 degrees of freedom which are the 2 (tranverse) photons.
Generalizing  the action to any space-time is straightforward. 

\medskip

In order to mimic the construction of minimally modified theories of gravity, we relax some of the conditions that make Maxwell theory unique. 
In minimally modified gravity theories, one breaks the full space-time diffeomorphism invariance and keep only symmetry under three dimensional
diffeomorphisms. In the case of Maxwell theory,  there is only the one-dimensional local symmetry group $U(1)$ that we choose to break, and then there
is no remaining local symmetry in the theory. However, to be  close to the gravity case, we also decide to break the global Lorentz symmetry keeping
only the invariance under the subgroup of rotations $SO(3)$ that leaves $A_0$ invariant. In that sense, $A_0$ is similar to the lapse function 
in the context of Maxwell theory.
As a consequence,
we look for theories whose action is of the form
\bea
\label{general lagrangian}
S[A_0,A_i] \; = \; \int d^4x \,  {\cal L}(A_0,\dot A_0,A_i,\dot A_i,\partial_i) \, ,
\eea
where ${\cal L}$ is the Lagrangian density. In other words, $\cal L$ is constructed from $A_0$, $A_i$, their first time derivatives and their space derivatives
at any order.

As we are going to see in a few lines, this theory propagates generically more than 2 degrees of freedom.
To find the conditions for the theory to propagate only 2 degrees of freedom, we perform a Hamiltonian analysis.
Hence, we start by introducting   the phase-space variables
\bea
\label{canvar}
\{A_\mu(x),P^\nu(y)\} \; = \; \delta_\mu^\nu \, \delta^3(x-y) \, .
\eea
If there is no constraints, the theory propagates 4 degrees of freedom. The presence of a primary constraint is then an obvious necessary condition for the theory to 
propagate only 2 (vectorial) degrees of freedom. The theory admits a primary constraint if its action is degenerate, i.e. the 4 dimensional Hessian matrix defined by
\bea
\label{Hessian}
\mathbb H^{\mu\nu} \equiv \frac{\partial^2 {\cal L}}{\partial \dot A_\mu \partial \dot A_\nu} \, ,
\eea
for $\mu,\nu \in \{0,1,2,3\}$ is not invertible. Furthermore, as we want vector modes to propagate, we add the condition that the submatrix $\mathbb H^{ij}$, for $i,j \in \{1,2,3\}$, 
is invertible. If this is the case, we can formally reformulate the Lagrangian density in \eqref{general lagrangian} as a function
\bea
{\cal L}(A_0,\dot A_0,A_i,\dot A_i,\partial_i) \; = \; {\cal F}(A_0,A_i,\dot A_i - \alpha_i \dot A_0 ,\partial_i) \, ,
\eea
where $\alpha_i$ depends on the connection $A_\mu$ and their spatial derivatives in general. In general (even when there is no coupling to external current)  
time derivatives of $A_0$  cannot be absorbed into a redefinition of $A_i$. But, for simplicity,  we assume that $A_0$ is not a dynamical variable as in the original Maxwell theory, 
and then it does not appear a priori with time derivatives in the action, which means that $\alpha_i=0$.
In that case, the theory possesses the simple primary constraint\footnote{The generalization to a non-zero $\alpha_i$ is immediate and the primary constraint is replaced by 
the combination ${\cal P} \equiv P^0 + \alpha_i P^i \approx 0$.}
\bea
\label{primary}
{\cal P} \equiv P^0 \approx 0 \, ,
\eea
where we recall that $\approx$ means weakly vanishing,

At this stage, there is no more primary constraint (which is a consequence of the fact that $\mathbb H^{ij}$ is not degenerate), and then one can (in principle) uniquely express (at least locally, on any open set of the phase space) the velocities $\dot A_i$ in terms of the momenta  $P^i$. As a consequence, one can construct (formally) the canonical and the total Hamiltonians, respectively given by
\bea
\label{totalH}
H \; = \; \int d^3x \, {\cal H}(A_\mu,P^i,\partial_i) \, , \qquad
H_{tot} \; = \; H +  \int d^3x \, \lambda \, P^0 \, ,
\eea
where $\lambda$ is a Lagrange multiplier which enforces the primary constraint. 

As we have already emphasized above, the relation between the Lagrangian and the canonical Hamiltonian is, in general, implicit. It can be made explicit in simple cases only, for free (quadratic) Lagrangians for instance. Furthermore, it will be much more convenient to find the conditions
for the theory to propagate (at most) 2 degrees of freedom in its Hamiltonian formulation than in its Lagrangian formulation. For all these reasons, we 
will construct modified Hamiltonian Maxwell theories, and in some cases, we will show how to recover the associated Lagrangian. 

\subsection{Killing the extra degrees of freedom}
\label{killing}
From now on, the starting point is the Hamiltonian \eqref{totalH} together with the primary constraint \eqref{primary}.
The stability under time evolution of the primary constraint leads  to a secondary constraint
\bea
\label{calS}
{\cal S} \; \equiv \; \{ {\cal P} \, , H \} \, = \, \frac{\partial \cal H}{\partial A_0} - \partial_i  \left(\frac{\partial  \cal H}{\partial (\partial_i A_0)}\right) 
+  \partial_i  \partial_j \left(\frac{\partial  \cal H}{\partial (\partial_i \partial_j A_0)}\right)  + \cdots \, \approx 0 \; ,
\eea
when $ \cal H$ depends explicitly on $A_0$. { In the particular case where $\cal H$ does not depend on $A_0$ (and on its spatial derivatives), 
then the Lagrangian itself does not depend on $A_0$ and the theory propagates 3 degrees of freedom. For this reason, we assume from now on that $\cal H$ depends on 
(the spatial derivatives of) $A_0$. To be more precise, we exclude the case where $\mathcal{H}$ depends on $A_0$ and its spatial derivatives only through a total spatial derivative.}

Even in that case, the theory could propagate
up to 3 degrees of freedom (if there is no more constraint and if the two constraints are second class).
To go further and to find the conditions on the Hamiltonian for the theory to propagate (at most) two degrees of freedom, we compute 
the Poisson bracket  between the primary and the secondary constraints,
\bea
\Delta (x,y) \; \equiv \; \{ \, {\cal S}(x) \, , \,  {\cal P}(y) \} \, ,
\eea
and one studies whether it (weakly) vanishes or not. Notice that we are using the shortened notations $F(x)=F(A_\mu(x),P^i(x),\partial_i)$ for any
function $F$ in the phase space. 

First, we study  the case where $\Delta$ is not weakly vanishing.  
There are no more constraints in the theory, and the pair ($\cal P$, $\cal S$) form a set of second class constraints. 
Hence, the theory propagates
$[(2\times 4) - 2]/2=3$ degrees of freedom, i.e. one more  than Maxwell theory. The extra degree of freedom is the longitudinal mode which comes
with the usual two polarizations of the graviton. 

Now, we study the more interesting case where $\Delta$ is weakly vanishing. 
The number of degrees of freedom depends on whether the bracket $\Omega(x,y) \equiv \{ {\cal S}(x)  , { \cal H}(y)\}$ is vanishing or not.
If $\Omega$ is weakly vanishing, the theory has no more constraints, the pair $({\cal P},{\cal S})$ forms a set of first class constraints, 
which means that there is a ``hidden" local symmetry in the theory. Furthermore, the theory propagates $[(2\times 4)-(2\times 2)]/2=2$ degrees of freedom, as in the Maxwell theory. 
If  $\Omega$ is not weakly vanishing, there is a tertiary constraint ${\cal T}$,  {but this may be not enough to insure that the theory propagates
2 degrees of freedom only. If one of the three constraints is first class (which is necessary the case if all the constraints are local), then the theory admits an extra symmetry and only
2 degrees of freedom. If this is not the case, one needs the presence of an extra quaternary constraint which would definitively imply that there is strictly less than $2$ degrees of freedom. }

As a consequence, in any cases, we see that a necessary condition for the theory to propagate 2 or less degrees of freedom is that 
\bea
\label{conditionfor2DOF}
\{ {\cal P}(x) \, , \, \{  {\cal P}(y), H \}\} \; \approx \; 0 \, ,
\eea
i.e. it vanishes up to terms proportional to ${\cal S}$. 
Let us make this condition more 
explicit, and show that it necessarily implies that ${\cal S} \equiv \{  {\cal P}, H_0 \}$ does not depend on $A_0$. For that, let us assume
the reverse is true, and then ${\cal S}$ is supposed to depend at least on $A_0$ or on one of its spatial derivatives.
Hence, the constraint  ${\cal S}(A_0,\partial_i A_0, \cdots)\approx 0$ can be viewed as a differential equation 
that we can solve for $A_0$ (with appropriate boundary conditions) in terms of the remaining phase space variables, at least formally.
In that case,
the secondary constraint can be (locally) replaced by the equivalent constraint
\bea
\tilde{\cal S} \; \equiv \; A_0 - {\cal A}_0(A_i,P^i,\partial_i) \, \approx \, 0 \, ,
\eea
where  ${\cal A}_0$ is the explicit solution for $A_0$. As a consequence, the new bracket between the constraints 
$ \{ \tilde{\cal S}(x)  ,   {\cal P}(y) \}  = \delta (x-y)$ is clearly non-vanishing, and then the theory propagates 3 degrees of freedom, which contradicts the initial assumption. As a consequence, 
the condition \eqref{conditionfor2DOF} is (locally) equivalent to the condition that $\cal S$ can be written as
\bea
{\cal S} \; = \; \nu(A_0) \, {\cal H}_0(A_i,P^i,\partial_i) \, ,
\eea
where ${\cal H}_0$ 
does not depend neither on $A_0$ nor on its derivatives, and $\nu$ is an arbitrary non-vanishing function of $A_0$, say positive. 
Hence, the Hamiltonian density takes necessarily the form {(up to a total spatial derivative)}
\bea
{\cal H} = {\cal V} + N(A_0){\cal H}_0\, ,
\eea
where ${\cal H}_0$ and ${\cal V}$ depends on $A_i, P^i$ and their spatial derivatives only. The function $N$ is an integral of $\nu$, and
then it is an increasing function of $A_0$ (as $\nu$ is supposed to be positive). 
Furthermore a simple canonical transformation allows us to fix (locally) $N(A_0)=A_0$ without loss of generality. 

\subsection{Complete Hamiltonian description}
To summarize, we found that any Hamiltonian theory which satisfies the necessary condition 
\eqref{conditionfor2DOF} is defined (up to a canonical transformation) by a phase space parametrized by the 4 
pairs of canonical variables \eqref{canvar} whose dynamics is governed
by a Hamiltonian of the form
\bea
\label{Ham2dof}
H \; = \; \int d^3x \, \left[  {\cal V}(A_i,P^i,\partial_i) + A_0 \, {\cal H}_0(A_i,P^i,\partial_i) \right] \, ,
\eea
together with the primary constraint ${\cal P} \approx 0$ \eqref{primary}. 
Hence, the secondary constraint is now simply given by
\bea
{\cal S} \; \equiv \; {\cal H}_0(A_i,P^i,\partial_i) \; \approx \; 0 \, .
\eea
The existence of this constraint implies immediately that the constraint ${\cal P} \approx 0$ is in fact first class, and it corresponds to the 
(on-shell) invariance of the theory under the arbitrary shift,
\bea
\label{shiftsymA0}
A_0 \mapsto A_0 + u \, ,
\eea 
of the non-dynamical variable $A_0$, by an arbitrary function $u(x)$.

Requiring conservation under time evolution of the secondary constraint leads to the condition
\bea
\label{condHH}
\int d^3y \, \left( \{ {\cal H}_0(x), {\cal H}_0(y) \} A_0(y) + \{ {\cal H}_0(x), {\cal V}(y)\}\right) \; \approx \; 0 \, ,
\eea
{whose resolution depends on the properties of $\Delta(x,y) \equiv \{ {\cal H}_0(x), {\cal H}_0(y) \} $ viewed as an operator acting on the space of functions  Fun$(\mathbb R^3)$
by integration.  When $\Delta$ is invertible, the condition \eqref{condHH} fixes completely the Lagrange multiplier $A_0$ in terms of the phase space variables.
Furthermore, in that case,  $\Delta(x,y)$ is necessary not scalar\footnote{A two-point distribution $F(x,y)$ is scalar if and only if $F(x,y)=F(x,0) \delta(x-y)$ where $\delta$ is the Dirac distribution.} (it involves derivatives of Dirac distributions) and the constraint on $A_0$ is in fact a partial differential equation
which would need appropriate boundary conditions to be explicitly resolved. There is no quaternary constraint as the time evolution of ${\cal T} \approx 0$ fixes completely 
the Lagrange multiplier. Then the theory admits three secondary constraints with a non-scalar Dirac matrix and a non-scalar Poisson bracket between $\cal P$ and $\cal T$ in particular. As a consequence, the theory is not well-posed. }

{The case where $\Delta$ is a non (weakly) vanishing operator with a non-trivial kernel is much more complicated to study. To understand this situation, it is convenient to decompose the space of functions on which $\Delta$ acts as the direct sum Fun$(\mathbb R^3)$=Im($\Delta$) $\oplus$ Ker($\Delta$) where Im($\Delta$) and Ker($\Delta$) are respectively the image and the kernel of $\Delta$. Hence, the condition \eqref{condHH} not only fixes the component of $A_0$ in  Im($\Delta$) but also can produce a new (quaternary) constraint obtained by projecting \eqref{condHH} into Ker($\Delta$). The new constraint may be non-scalar,  the general Dirac analysis appears to be very subtil, and it should be done on a case-by-case basis.  For that reason, we will exclusively consider the simpler case where $\Delta$ is weakly
vanishing:}
\bea
\label{comSS}
\{ {\cal H}_0(x), {\cal H}_0(y)\} \; \approx \; 0 \, .
\eea
If this is the case, the conservation of the secondary constraint under time evolution leads either to a tertiary constraint
\bea
{\cal T}(x) \; \equiv \; \{ {\cal H}_0(x) \, , \, \int d^3y \, {\cal V}(y)\} \, ,
\eea
or to no new constraint if ${\cal T} $ is itself weakly vanishing. In any of these two cases, the theory propagates 2 degrees of freedom or less.
\begin{itemize}
\item {Case where ${\cal T} \approx 0$ is automatically satisfied}. The theory admits 2 first class constraints ${\cal P}  \approx 0$ and 
${\cal H}_0  \approx 0$. The constraint $\cal P$ is associated to the (on-shell) symmetry described above \eqref{shiftsymA0}, and the constraint ${\cal S}$ generates a gauge symmetry acting on the phase space variables $(A_i,P^i)$, exactly as in Maxwell
theory. As a result the theory propagates $[(2\times 4)-(2\times 2)]/2=2$ degrees of freedom.
\item {Case where ${\cal T}  \approx 0$ is a new constraint, and $\cal T$ does not commute with ${\cal H}_0$}. The Dirac analysis stops here with
one first class constraint ${\cal P}  \approx 0$ and two second class constraints ${\cal H}_0  \approx 0$, $ {\cal T}   \approx0$, which lead
 to $[(2 \times 4) - (2+1+1)]/2=2$ degrees of freedom.
 \item   {Case where ${\cal T}  \approx 0$ is a new constraint, and $\cal T$ commutes with ${\cal H}_0$}. Either the Dirac analysis continues producing
 constraints, or $\cal T$ and ${\cal H}_0$ are first class. In any case, the theory propagates 1 or 0 degree of freedom.
\end{itemize}
{As a conclusion, any deformation of Maxwell theory which breaks the $U(1)$ symmetry, which is invariant under the global $SO(3)$ group that leaves
$A_0$ invariant and which propagates at most 2 degrees of freedom has necessarily a Hamiltonian of the form \eqref{Ham2dof}.  Furthermore, the condition \eqref{comSS}
is sufficient to insure that the theory propagates at most 2 degrees of freedom, but it has not been rigorously proven that it is also necessary because the theory admits a quaternary 
(eventually non-local) constraint  when $\Delta$ is non-vanishing with a non-trivial kernel.}

\subsection{Example: quadratic theories}
Let us illustrate the previous analysis with a simple example. We consider a Hamiltonian which is, at most, quadratic in the phase space
variables $(A_0,A_i,P^i)$. 

\subsubsection{General Hamiltonian analysis}
Furthermore, we assume that the Hamiltonian can be written in terms of the fields and their first order (spatial) derivatives only. In that case ${{\cal H}_0}$ is linear in $(A_i,P^i)$ whereas
${{\cal V}}$ is quadratic in $(A_i,P^i)$. Hence, these two functions can be written as
\bea
{\cal V} & = &  \alpha_1 \, A^2+ \alpha_2 \, P^2+ \alpha_3 \, (AP) + \alpha_4 \, (\partial A)^2 + \alpha_5 \, (\partial P)^2 + \alpha_6 \, (\partial A)(\partial P) \nonumber \\
&-&  \alpha_7 \, \partial_j A_i \partial^j A^i - \alpha_8 \, \partial_j P_i \partial^j P^i - \alpha_9 \, \partial_j A_i \partial^j P^i \, ,\\
{\cal H}_0 & = & \beta_1 \, \partial  A + \beta_2 \, \partial  P\, , 
\label{quadraticH1}
\eea
where $\alpha_I$ and $\beta_I$ are constant, and we used the shortened notations
\bea
\partial X \equiv \partial_i X^i \, , \quad
XY \equiv X_i Y^i \, , \quad
X^2\equiv X_i X^i \, ,
\eea 
for $X$ being $A$ or $P$. Notice that indices are lowered and raised with the flat metric $\delta_{ij}$ and its inverse $\delta^{ij}$.
As ${\cal H}_0$ trivially satisfies the condition \eqref{comSS}, the theory propagates at most 2 degrees of freedom for any values of
the coefficients $\alpha_I$ and $\beta_I$.

Let us study these theories in details. First, using canonical transformations, we can simplify the shape of the Hamiltonian. 
Indeed, canonical transformations (with no explicit time dependency) which preserves quadratic and first order Hamiltonians 
are of the form
\bea
A \; \longmapsto \; x A + y P \, , \qquad
P \; \longmapsto \; z A + w P \, , \qquad xw-yz=1 \, .
\eea
Hence, (when $\beta_2 \neq 0$) one can find a canonical transformation such that 
\bea
{\cal V} & = &  \alpha_1 \, A^2+ \alpha_2 \, P^2+ \alpha_4 \, (\partial A)^2 + \alpha_5 \, (\partial P)^2 + \alpha_6 \, (\partial A)(\partial P) \nonumber \\
&-& \alpha_7 \, \partial_j A_i \partial^j A^i - \alpha_8 \, \partial_j P_i \partial^j P^i - \alpha_9 \, \partial_j A_i \partial^j P^i \, ,\\
{\cal H}_0 & = & - \partial  P\, , 
\label{quadraticH}
\eea
which corresponds to taking $\alpha_3=0$, $\beta_1=0$ and $\beta_2=-1$ in the general expression \eqref{quadraticH1}. As a consequence,
the expression of the constraint ${\cal H}_0 \approx 0$ has exactly the same form as in Maxwell theory, and then, one can fix $\alpha_5=0$
without loss of generality (by a redefinition of the Lagrange multiplier $A_0$). Notice that, even though the constraint ${\cal H}_0  \approx 0$ 
is the same as in Maxwell theory, it is not necessarily first class. This can be easily seen if one re-expresses the total Hamiltonian
as follows
\bea
\label{Hamiltoniancurv1}
H & = & \int d^3x \,[- A_0 \partial P + \alpha_2 P^2 - \frac{1}{2} \alpha_4 F_{ij} F^{ij} + \alpha_6 F_{ij} \partial^j P^i 
+ (\alpha_9 - \alpha_6) P_i \Delta A^i ] \nonumber \\
&& \qquad + [ \alpha_1 A^2 + (\alpha_7-\alpha_4) A_i \Delta A^i ] \, ,
\eea
where $F_{\mu\nu}$ is the curvature of the connection \eqref{curvature}. 
The first line in \eqref{Hamiltoniancurv1} is invariant under the $U(1)$ gauge symmetry $\delta_\varepsilon A_i = \partial_i \varepsilon$. 
The second line is clearly not, which makes the constraint second class, and,
from its expression, we see that the conditions for the theory to be $U(1)$ gauge invariant are immediately given by
$\alpha_1=0$ and $\alpha_7=\alpha_4$.

For the moment, let us complete the Hamiltonian analysis. Using the notations of section \ref{killing}, the secondary constraint is
${\cal S} \equiv \partial P$. To compute the remaining constraints, it is convenient to first write the equations of motion:
\bea
\dot A_i & =  & \{ A_i,H_0 \}  \; \approx \;  \partial_i A_0 + 2 \alpha_2 P_i  - \alpha_6 \partial_i (\partial A)
+2\alpha_8 \Delta P_i + \alpha_9 \Delta A_i \,  , \label{HamEOMA}\\
\dot P_i & =  & \{ P_i,H_0 \}  \; \approx \; -2\alpha_1 A_i + 2 \alpha_4 \partial_i (\partial A) -2 \alpha_7 \Delta A_i - \alpha_9 \Delta P_i \, .  \label{HamEOMB}
\eea
The tertiary constraint is obtained from the requirement that ${\cal S}\approx 0$ has to be weakly 
conserved under time evolution, which means that
\bea
\dot{\cal S} \approx 0 \; \Longleftrightarrow\;  \partial_i \dot P^i \approx 0 \; \Longleftrightarrow\; 
{\cal T} \equiv [\alpha_1 + (\alpha_7 - \alpha_4) \Delta ] (\partial A) \approx 0 \; .
\eea
Using suitable boundary conditions, one can replace the constraints $\cal T$  by the condition
\bea
\partial A \approx 0 \, ,
\eea
except if $\alpha_1 =0$ and $\alpha_4=\alpha_7$, in which case the constraint $\cal S$ is first class, as we have already seen previously.
Clearly, the constraints $\cal S$ and $\cal T$ do not commute, and then the conservation of $\cal T$ under time evolution does not lead to any new constraints. As a conclusion, the  Dirac analysis of the theory closes with one first class constraint ${\cal P} \approx 0$ and the two second class constraints ${\cal S}\approx 0$ and ${\cal T} \approx 0$. This leads to 2 degrees of freedom, as expected. 

\subsubsection{Lagrangian}
{Let us focus on the case with $\alpha_1\neq 0$ or $\alpha_4\neq \alpha_7$, in which the theory has one first-class constraint and two second-class constraints.}
The first class constraint allows us to choose a gauge where $A_0=0$. In this gauge, the equations
of motion \eqref{HamEOMA} and \eqref{HamEOMB}  simplify into  
\bea
\dot A_i & = & 2 (\alpha_2   +  \alpha_8 \Delta) P_i + \alpha_9 \Delta A_i \; , \\
-\dot P_i & = &  2 (\alpha_1  +  \alpha_7 \Delta) A_i + \alpha_9 \Delta P_i \, ,
\eea
with the constraints that $\partial P = 0 = \partial A$, which means that both vectors are transverse. From this, we immediately see
that the theory admits 2 degrees of freedom only which are governed, after decoupling the previous system, by the  equation
\bea
\label{eomforA}
-\ddot A_i + \alpha_9 (\Delta \dot A_i - \dot A_i + \alpha_9 \Delta A) - 4(\alpha_2 + \alpha_8 \Delta)(\alpha_1 + \alpha_7 \Delta) A_i = 0 \, .
\eea
Notice that this equation is second order in time but higher order (up to fourth order) in space. This ensures that the theory is healthy and does not propagate 
Ostrogradski ghosts. {Notice that the presence of higher space derivatives in the equations of motion could mean the existence of generalized instantaneous mode
(or shadowy modes) which would appear in a ``covariantization'' of the theory similar to what happens in scalar-tensor theories \cite{DeFelice:2018mkq}}.

\medskip

It is also instructive to compute the Lagrangian and study some of its properties. As the Hamiltonian is quadratic, the associated Lagrangian   is easily obtained from the Legendre transformation 
\bea
{L}[A_\mu] \; = \; \int d^4x \, \left(P \dot{A} - {\cal V} + A_0 \partial P \right) \, ,
\eea
where the momenta $P^i$ are expressed in terms of the velocities $\dot A_i$ solving the equation of motion \eqref{HamEOMA}.
Formally the momenta variables are given by
\bea
P_i & = & \frac{1}{2}(\alpha_2 + \alpha_8 \Delta)^{-1} [\dot A_i - \partial_i A_0 + \alpha_6 \partial_i (\partial A) - \alpha_9 \Delta A_i] \nonumber \\
& = &  \frac{1}{2}(\alpha_2 + \alpha_8 \Delta)^{-1} \left[F_{0i} + \alpha_6 \partial^j F_{ij} + (\alpha_6 - \alpha_9) \Delta A_i \right] \, ,
\eea
which, to be defined, needs suitable spatial boundary conditions. 

First, we immediately remark that a non-vanishing $\alpha_8$ coefficient in the Hamiltonian makes the Lagrangian (spatially) non-local. In general, any terms which involve spatial derivatives of the momenta in the Hamiltonian will produce non-local terms in the Lagrangian, even though we started from a local Hamiltonian. For simplicity, we restrict the analysis to local Lagrangians, which, in this case, implies 
$\alpha_8=0$.  Notice that $P_i$ is not $U(1)$ gauge invariant when $\alpha_6 \neq \alpha_9$. 
The calculation of the Lagrangian is now immediate and shows that it contains higher spatial derivatives but not higher time derivatives as expected. This is obviously consistent
with the equation of motion for the vector field $A_i$ \eqref{eomforA}.

\subsubsection{Modified gauge invariant Maxwell theories}

To finish with this example, let us consider quadratic theories which are gauge invariant, i.e. 
$\alpha_1=\alpha_7-\alpha_4=0$. {For simplicity, we assume $\alpha_9-\alpha_6=0$ as well as $\alpha_8=0$.} In that case, ${\cal H}_0 \approx 0$ is first class, and the full connection transforms as expected according to $A_\mu \mapsto A_\mu + \partial_\mu \theta$ where $\theta$ is an arbitrary function, under the symmetry. The infinitesimal transformation law of $A_i$ comes from the Poisson action of  ${\cal H}_0$,
\bea
\delta_\varepsilon A_i \; = \; \{A_i , \int d^3x \, \varepsilon(x) {\cal H}_0 \} \, .
\eea
{The transformation law for $A_0$ under gauge transformations can be seen from the
gauge invariance of the full (covariant) Lagrangian}.  In that context, the canonical Hamiltonian is simply given by
\bea
\label{Hamiltoniancurv}
H \; = \; \int d^3x \, \left[- A_0 \partial P + \alpha_2 P^2 - \frac{1}{2} \alpha_4 F_{ij} F^{ij} + \alpha_6 F_{ij} \partial^j P^i \right] \, ,
\eea
and, after some calculations, one finds that the action is given by
\bea
S[A_0,A_i]\; = \; \frac{1}{2} \int d^4x \; \left[ -\frac{1}{2\alpha_2} F_{0i} F^{0i} + \alpha_4 F_{ij} F^{ij} - \frac{\alpha_6^2}{2\alpha_2} 
(\partial_j F^{ij})^2 \right] \, .
\eea 
As expected, the action is not Lorentz invariant, it contains spatial higher derivatives terms and its equations of motion are given by
\bea
\partial_i F^{0i}=0 \, , \qquad
\partial_0 F^{0i} + 4 \alpha_2 \alpha_4 \partial_j F^{ij} + \frac{\alpha_6^2}{2} \Delta \partial_j F^{ij}=0 \, ,
\eea
which we can compare to standard Maxwell equations $\partial_\mu F^{\mu\nu}=0$. Using the usual definitions of the electromagnetism
field $E^i \equiv F^{0i}$ and $B^i \equiv \varepsilon^{ijk} F_{jk}$, we obtain the following modified Maxwell equations in the vacuum,
\bea
\text{div} \vec{E} = 0 \, , \quad
(1+ \mu \Delta) \vec{\text{rot}} \vec{B} = \lambda \frac{\partial \vec{E}}{\partial t} \, ,
\eea
where $\mu=\alpha_6^2/(8 \alpha_2 \alpha_4)$ and $\lambda = -1/(4\alpha_2\alpha_4)$. The equations $\text{div} \vec{B}=0$
and $\vec{\text{rot}}\vec{E}+ \partial \vec{B}/\partial t=0$ which are equivalent to the existence of the gauge field $A_\mu$ are
obviously unchanged. Hence, the propagation equations become
\bea
\Delta \vec{V} - \lambda \frac{\partial^2 \vec{V}}{\partial t^2} + \mu \Delta^2 \vec{V} \; = \; \vec{0} \, ,
\eea
where $\vec{V}$ can be either $\vec{E}$ or $\vec{B}$.
It is obvious that $\lambda$ parametrizes the deviation to the speed of light and $\mu$ parametrizes higher derivatives
deviations. As the theory is still linear and $\lambda$ is constant, we can fix it to $\lambda=1$ by a rescaling of the time variable, {provided
that $\lambda$ is neither vanishing nor divergent}.

To close the analysis of this example, let us make a couple of remarks.
First, It is easy to generalize our analysis to cases where higher derivatives have an order higher than two, including in $H$ terms with higher than 2 spatial derivatives of the fields $A_\mu$. Introducing higher derivatives of the momenta variables would produce non local actions. 

Second, as we briefly discussed below \eqref{Hessian}, one could have started with a dynamical $A_0$ variables in the Hamiltonian framework. For that, one
would have replaced the primary constraint by a more general constraint ${\cal P}(P^0,P^i) \approx 0$ 
which would mix all the components of the momenta. The analysis would be similar to what we have done. Another way to make $A_0$
dynamical would be to the consider ``disformal-like" transformations on the connection which preserve the quadratic form:
\bea
A_0 \mapsto A_0 + x \partial_i A^i \, , \quad 
A_i \mapsto A_i + y \partial_i A_0 \, ,
\eea
where $x$ and $y$ are constant.

\section{Generalization to gravity}
\label{gravity}
In this section, we adapt the previous construction to gravity, and we  construct a large class of minimally modified gravity theories  from
the Hamiltonian point of view. We first find (sufficient) conditions on the Hamiltonian for the theory to propagate at most two tensorial degrees of freedom.
Then, we illustrate our construction with examples. In particular, we will exhibit a new interesting class of minimally modified gravities, dubbed 
$f({\cal H})$ theories. 

We start with the ADM parametrization of the metric in terms of the lapse function $N$, the shift vector $N^i$ and the induced spatial 
$h_{ij}$, as it was recalled in the introduction \eqref{ADM}.

\subsection{The modified phase space}
The phase space is parametrized by
the usual ten pairs of canonical variables
\bea
&&\{h_{ij}(x),\pi^{kl}(y) \} = \delta_{ij}^{kl} \, \delta(x-y) \, , \\
&&\{N^i(x),\pi_j(y)\}=\delta^i_j \, \delta(x-y) \, , \\
&&\{N(x),\pi_N(y)\}=\delta(x-y) \, .
\eea
We want to construct a Hamiltonian in this phase space 
which satisfies the properties of minimally modified gravity, i.e.
\begin{itemize}
\item It is invariant under space-like diffeomorphisms;
\item  It propagates only 2 tensorial degrees of freedom (or less);
\item The lapse and the shift are non-dynamical.
\end{itemize}
Notice that the last requirement is not necessary, and one can relax the condition that the lapse function is not dynamical at the price to add a degeneracy condition
as it is done in the context of DHOST theories \cite{Langlois:2015cwa}.   
Another way to make the lapse function dynamical would be to perform a disformal transformation on the metric variables.
For simplicity, we will consider only the case where $N$ is not dynamical.

The invariance under space-like diffeomorphisms implies immediately that the canonical Hamiltonian takes the form
\bea
\label{MMGHamiltonian}
H = \int d^3x \, \sqrt{h} \, \left[ {\cal H}(h_{ij},\pi^{ij},N,\nabla_i) + N^i {\cal V}_i \right] \, ,
\eea 
where ${\cal V}_i \equiv -2\nabla^j (\pi_{ij}/\sqrt{h})$ is the usual vectorial constraint of gravity, and $\cal H$ is a priori an arbitrary scalar.
At this stage, with no restriction on the function $\cal H$, it is straightforward to see that the theory generically propagates 3 degrees of freedom.

Following what has been done for Maxwell theory in the previous section, we can immediately show that a necessary condition 
(up to a redefinition of the lapse function by a canonical transformation) for the theory to propagate (up to) 2 degrees of freedom is that
\bea
{\cal H} \; = \; {\cal V} + N \, {\cal H}_0 \, ,
\eea
where ${\cal H}_0$ and ${\cal V}$ are three-dimensional scalar which depend on $h_{ij}$, $\pi^{ij}$ and their covariant spatial derivatives only.  
The fact that there are scalars insures that they commute with the vectorial constraint. Hence, the conservation under time evolution of the
constraints $\pi_N \approx 0$ and $\pi_i \approx 0$ creates respectively  the constraints
\bea
{\cal H}_0 \; \approx \; 0 \, , \qquad {\cal V}_i \; \approx \; 0 \, .
\eea
By construction, the vectorial constraints ${\cal V}_i  \approx 0$, together with $\pi_i \approx 0$, are necessarily first class.

Then, requiring that the theory has enough constraints to kill the extra degrees of freedom implies, as in the vector case, {leads to the condition
that $\{ {\cal H}_0(x)  , {\cal H}_0(y) \}$ has necessarily a non-trivial kernel (see \eqref{condHH} and the paragraph below). A sufficient condition is that}
\bea
\label{H1condition}
\{ {\cal H}_0(x) \, , \, {\cal H}_0(y) \} \; \approx \; 0 \, ,
\eea
{and we restrict our analysis to that case only}
where the conservation under time evolution of ${\cal H}_0 \approx  0$ implies the condition
\bea
\{ {\cal H}_0(x) \, , \, {\cal V}(y) \} \; \approx \; 0 \, .
\eea
If this condition is trivially (weakly) satisfied, then there is no tertiary constraints in the theory. The constraints ${\cal H}_0 \approx 0$ and $\pi_N \approx 0$ are also first class,
 and the theory propagates $[10 \times 2 -(3\times 2 + 3 \times 2 + 1 \times 2 + 1 \times 2)]/2=2$ degrees of freedom, as in Einstein theory. 
 Furthermore, in that case, the theory admits an extra symmetry in addition to three-dimensional diffeomorphisms. 
 
 If, on the contrary, the condition \eqref{H1condition} is not trivially satisfied, then the  theory admits a new constraint which is
\bea
{\cal T}(x) \; \equiv \; \{ {\cal H}_0(x) \, , H \} \; \approx \; 0 \; .
\eea
 The existence of this new constraint is sufficient to conclude that  the theory  propagates at most 2 degrees of freedom. Indeed, as the constraint
 $\pi_N \approx 0$ is necessarily first class (because the theory is invariant by any redefinition of the lapse), the theory admits 7 first class constraints
 in addition to  the two constraints ${\cal H}_0 \approx 0$ and ${\cal T} \approx 0$, which implies immediately that the theory propagates 2 or less  
  degrees of freedom. It has exactly 2 degrees of freedom if ${\cal H}_0$ and ${\cal T}$ are second class, and no degrees of freedom if they are first class.

\medskip

As a conclusion,
{the following Hamiltonian satisfies the three
conditions recalled at the beginning of this section and thus defines
a class of minimally modified theories of gravity:}
\begin{eqnarray}
&&H = \int d^3x \, \sqrt{h} \, \left[ {\cal V}(h_{ij},\pi^{ij},\nabla_i) + N  {\cal H}_0(h_{ij},\pi^{ij},\nabla_i)-2 N^i \nabla^j \left(\frac{\pi_{ij}}{\sqrt{h}}\right)\right] \, , \label{HMMG} \\
&&\text{with} \qquad \{ {\cal H}_0(x) \, , \, {\cal H}_0(y) \} \; \approx \; 0 \, . \label{selfcommutation}
\end{eqnarray}
{In that case, the function ${\cal V}$ is totally free. 
Notice that, as $N$ and $N^i$ are not dynamical, the Hamiltonian comes with the primary constraints $\pi_i \approx 0$ and $\pi_N \approx 0$ which are
first class. They are associated to the invariance of the theory under arbitrary redefinitions of the lapse and shift. }

\subsection{Simple examples: ${\cal H}_0$ is the Hamiltonian constraint and ${\cal V}$ is polynomial in $\pi$}
To illustrate the previous general construction, let us consider the simple example defined by
\bea
\label{Hgr}
{\cal H}_0 \; = \; \frac{1}{{\vert h \vert }} \left( \pi_{ij} \pi^{ij} - \frac{1}{2} \pi^2\right) -  R \, , \qquad
{\cal V} \; = \; \lambda \, \pi \, -  \mu \, \sqrt{\vert h \vert } \, ,
\eea
where $\lambda$ and $\mu$ are constant, and $R$ is the three-dimensional curvature. We notice that, as ${\cal H}_0$ is the Hamiltonian constraint of gravity, it trivially satisfies the condition
\eqref{H1condition}. In fact, if we fix ${\cal H}_0$ to this expression, one could have chosen any arbitrary function for ${\cal V}$ but  for simplicity we make the choice above. 

With this example, one can easily compute the explicit action which is given by
\bea
\label{actionmodel}
S \; = \; \int d^4x \, N \sqrt{h} \, \left[ 
K_{ij}K^{ij}-K^2+R + \lambda \left( \frac{K}{N} - \frac{3\lambda}{4 N^2} \right) + \frac{\mu}{N} 
\right] \, ,
\eea
where $K_{ij}$ is the usual extrinsic curvature
\bea
K_{ij} = \frac{1}{2N} \left( \dot h_{ij} - \nabla_i N_j - \nabla_j N_i\right) \, ,
\eea
and $K \equiv K_i^i$ is its trace.

Let us remark that the change of variable
\bea
K_{ij} \; \equiv \overline{K}_{ij} + \frac{\lambda}{2N} h_{ij} \, ,
\eea
allows to see that the previous action \eqref{actionmodel} takes exactly the same form as the general relativity action
\bea
S \; = \;  \int d^4x \, N \sqrt{h}  \left( \overline{K}_{ij} \overline{K}^{ij}-\overline{K}^2+R + \frac{\mu}{N}\right) \,,
\eea
up to the $\mu$-term.
However, as $\overline{K}_{ij}$ cannot be interpreted as the extrinsic curvature of a metric, the theory is not equivalent to general relativity.
To illustrate the difference between the modified theory and general relativity, let us now make the following time dependent change of variable on the metric components
\bea
\hat{h}_{ij} \; \equiv \; e^{-\lambda t} h_{ij} \, , \quad
\hat{N}_i \; \equiv \; e^{-\lambda t} N_i \, , \quad
\hat{N} \; \equiv \; e^{-\lambda t/2} N \, .
\eea
Hence, the action takes the form 
\bea
S \; = \;  \int d^4x \, \hat{N} \sqrt{\hat{h}} 
\left(\hat{K}_{ij}\hat{K}^{ij}-\hat{K}^2+e^{-\lambda t}\hat R + e^{3 \lambda t} \frac{\mu}{\hat{N}}\right) \,\eea
which makes  obvious that the theory propagates only 2 tensorial degrees of freedom because the modification affects only terms with spatial derivatives in the action. 

To finish with this example, let us remark that the action \eqref{actionmodel}  can easily be made covariance  introducing, as usual, a scalar field $\phi$ whose gradient is
orthogonal to the space-like hyper-surfaces. Using the results of \cite{Langlois:2017mxy}, one obtains
\bea
S[g_{\mu\nu},\phi] \; = \;  \int d^4x \, \sqrt{\vert g \vert} \left[ {\cal R} - \frac{\lambda}{2} \ln(X^2) \Box \phi + \frac{3\lambda^2}{2} X + 2 \mu \sqrt{-X}\right] \, . 
\eea
From this action, it is clearly not obvious that only two gravitational degrees of freedom are propagating. But the theory belongs to the class of ``cuscuton"
theories \cite{Afshordi:2006ad,Iyonaga:2018vnu}.

\medskip

A more interesting example would be to assume that ${\cal V}$ is a scalar quadratic in $\pi_{ij}$, in which case, it can be written as
\bea
{\cal V} \; = \; \frac{1}{{\vert h \vert}}\left( \lambda_1 \pi^{ij}\pi_{ij} -\frac{ \lambda_2}{2} \pi^2 \right) \, ,
\eea
where $\lambda_1$ and $\lambda_2$ are constant. 

Using the results of the Hamiltonian analysis of DHOST theories \cite{Langlois:2015skt}, 
we see that such a Hamiltonian can be obtained from a DHOST theory in the unitary gauge with a k-essence term,
a generalized cubic galileon term and a quadratic DHOST term with
\bea
\frac{a_1}{N^2}+1 = \frac{1}{N + \lambda_1} \, , \quad
 \frac{a_2}{N^2}-1 = \frac{N + \lambda_2}{(N+\lambda_1)(2\lambda_1 - 3 \lambda_2 - N)} \, ,
\eea
in the unitary gauge. 
We notice that the theory belongs to (the safe) class I only if $a_1+a_2=0$, which implies that $\lambda_1=\lambda_2$. Otherwise, perturbations about any cosmological background develop gradient instabilities. 
Furthermore, all these theories belong by definition to the class of extended cuscuton \cite{Afshordi:2006ad,Iyonaga:2018vnu}.

\subsection{A new class of theories: $f({\cal H})$ theories}
In this section, we introduce a new interesting class of minimally modified theories of gravity. To explain the  construction of this class, we first recall that a 
Hamiltonian of the form \eqref{MMGHamiltonian} corresponds to a theory with (up to) two tensorial modes only if the ``modified" Hamiltonian constraint ${\cal H}_0$
commutes with itself \eqref{selfcommutation}. The function $\cal V$ is a priori free, but to have a modified theory which is very close to general relativity, we make the choice ${\cal V}=0$.

 In order for the theory to propagate gravitational wave, it is necessary that ${\cal H}_0$ contains both $K_{ij}$ terms and  three dimensional curvature terms (like
the  Ricci scalar $R$) as in the expression of the Hamiltonian constraint of general relativity \eqref{Hgr}.  The presence of such terms makes difficult the problem of finding
an expression of ${\cal H}_0$ which is different from the usual Hamiltonian constraint. However, there is a simple modification that we can think about which is
\bea
{\cal H}_0 \; = \; f({\cal H}_{gr}) \qquad \text{with} \quad  {\cal H}_{gr} \equiv \frac{ 2\pi_{ij} \pi^{ij} -  \pi^2}{2{\vert h \vert }}  -  R \, ,
\eea
where $f$ is an arbitrary function. As ${\cal H}_{gr}$ is dimensionful, the function $f$ needs at least a mass scale to be defined, which could be the Planck mass and something else,
like the cosmological constant...

In that case, the modified Hamiltonian constraint satisfies the Poisson algebra
\bea
\{ {\cal H}_0(N_1),{\cal H}_0(N_2)\} \; = \; [f'({\cal H}_{gr})]^2  (N_1 \nabla_i N_2 - N_2 \nabla_i N_1) {\cal V}^i \, ,
\eea
which is, in general, non-linear. Obviously, the Poisson bracket weakly vanishes. Hence, we have found a new class of minimally modified theories of gravity
that we dub $f({\cal H})$ theories with reference to $f(R)$ theories. Contrary to $f(R)$ theories, $f({\cal H})$ theories do not propagate scalar modes, and the main
reason is  that the associated equations of motion remain second order. 

From a Legendre transformation, one can easily compute the corresponding action. Indeed, the equation of motion for $h_{ij}$ enables us to relate the momenta $\pi_{ij}$
to the extrinsic curvature $K_{ij}$ as follows
\bea
K_{ij} \; = \; \frac{f'({\cal H}_{gr})}{\sqrt{\vert h \vert}} \left( \pi_{ij} - \frac{1}{2} \pi h_{ij} \right) \, ,
\eea
from which we can implicitely  obtain $\pi_{ij}$ in terms of $K_{ij}$ because, in general, this equation is non-linear in $\pi_{ij}$. Nonetheless, one can compute the action
which, after a simple calculation, is given by
\bea
\label{minimalLag}
S[h_{ij},N,N^i] \; = \; \int d^4x \sqrt{\vert g \vert} \left[ \frac{2}{f'(C)}(K_{ij} K^{ij} - K^2) - f(C)  \right] \, ,
\eea
where $C$ is formally obtained by solving the equation
\bea
\label{constraintC}
C \; = \; \frac{K_{ij} K^{ij} - K^2}{[f'(C)]^2} - R \, .
\eea
In the case where $f(x)=x$, one immediately recovers the action of general relativity. However, any other choice for $f$ leads to a different theory which admits
a four dimensional symmetry algebra (the constraints satisfy a deformed diffeomorphisms algebra) and propagates only 
two tensorial modes. For instance, the choice $f(x)= x(1 - x/(2\Lambda))$ could be interesting for dark energy because the solutions of the deformed Hamiltonian constraint 
contain both a sector with no cosmological constant and a sector with a cosmological constant $\Lambda$. In fact, in any situation where $f(x)=0$ has a non-vanishing solution
$x_0$, there is in the theory an effective cosmological constant given by $x_0=2\Lambda$. For this reason, this new class of theories is very interesting and certainly deserves a 
deeper study.

\subsubsection{Hamilton equations}
We can easily compute the Hamilton equations of motion for any function ${\cal O}(h_{ij},\pi^{ij})$ in the phase space using the definition of the time derivative
\bea
\dot {\cal O}(x) \; = \; \{ {\cal O}(x)  \, , \, H \} \, .
\eea
The explicit form of the time derivative is easily obtained from the Hamilton equations of general relativity due to the fact that
\bea
 \{ {\cal O}(x)  \, , \, H \} & = & \int d^3y \, \sqrt{h(y)}  \, \left[ f'({\cal H}_{gr}(y)) \,N(y) \{ {\cal O}(x) \, , \, {\cal H}_{gr}(y)\} + N^i(y) \{ {\cal O}(x) \, , \, {\cal V}_i (y)\} \right] \nonumber \\
 &+&  \int d^3y \,  f({\cal H}_{gr}(y)) \, N(y) \{ {\cal O}(x) \, , \, \sqrt{h(y)}\} \, .
\eea
In the vacuum, the second line vanishes due to the constraint, but this is not the case in the presence of matter.

Applying this formula to the spatial metric $h_{ij}$ and its momenta $\pi^{ij}$ and using well-known results of Hamiltonian general relativity (see \cite{Poisson:2009pwt} for instance) 
leads immediately to the expressions
\bea
\dot{h}_{ij} & = & D_i N_j + D_j N_i + \frac{N f'}{\sqrt{h}} \left( 2 \pi_{ij} - \pi h_{ij}\right) \, , \label{Eqhij}\\
\dot{\pi}^{ij} & = & - \sqrt{h} N \left[f'  R^{ij} + \frac{1}{2} f h^{ij}\right]+ \sqrt{h} (D^i D^j - h^{ij} D^2) (N f') - D_k \left[ {2 N^{(i} \pi^{j)k} - N^k \pi^{ij}}\right] \nonumber \\
&&-\frac{N f'}{\sqrt{h}} \left[ 2 \pi_k^i \pi^{kj} - \pi \pi^{ij} - \left( \pi_{kl} \pi^{kl} - \frac{1}{2} \pi^2\right) h^{ij}\right] \,,\label{Eqpiij}
\eea
where $f$ and $f'$ are evaluated at ${\cal H}_{gr}$. Combining these two equations would allow us in principle to obtain the modified Einstein equations. 
To do so, one has to express $\pi_{ij}$ in terms of the extrinsic curvature $K_{ij}$ using the first equation
\bea
K_{ij} \; = \; \frac{f'({\cal H}_{gr})}{\sqrt{h}} \left( \pi_{ij} - \frac{1}{2} \pi h_{ij}\right) \, ,
\eea
and then one substitutes the obtained expression in the second equation of motion for $\pi_{ij}$.
When $f(x)=x$, we recover immediately
the Hamilton equations of general relativity using the Hamiltonian constraint ${\cal H}_{gr}=0$. 

{In the presence of matter, these equations have to be supplemented with source terms. However, describing explicitly how matter is coupled to the (modified) gravitational field
is subtle and has been analyzed in great details in \cite{Aoki:2018zcv,Aoki:2018brq}. A ``naive'' minimal coupling\footnote{If the matter is minimally coupled (with no derivative couplings) and is described
by a action $S_M$ associated to an energy-momentum tensor $T^{\mu\nu}$, then the equation for $h_{ij}$ \eqref{Eqhij} is unchanged, the deformed Hamiltonian constraint
becomes
\bea
f({\cal H}_{gr}) + 16\pi G_N \, N^2 T^{00} \; \approx \; 0 \, ,
\eea
and the equation for the momenta $\pi_{ij}$ contains a source term
\bea
\dot{\pi}^{ij} \; = \; \dot{\pi}_0^{ij} + \frac{\delta S_M}{\delta h_{ij}} \; = \;  \dot{\pi}_0^{ij} + 8\pi G_N \, {N} \sqrt{h} \left( T^{ij} - N^i N^j T^{00} \right) \, ,
\eea
where $ \dot{\pi}_0^{ij}$ is the expression of $\dot{\pi}^{ij}$ in vacuum given by \eqref{Eqpiij}. However, as we said, in general such a coupling leads to new propagating 
degrees of freedom in addition to the tensors and the matter.}  of the matter fields, for instance, would break the gauge invariance
generated by the first class constraint ${\cal H}_0$ which, as a consequence, would become second class. Therefore, in general, one extra mode (besides those of the matter field) appears in the phase space.  A consistent way to introduce the matter field is, before inclusion of the matter fields, to split the first class constraint into a pair of second class constraints by introducing a ``gauge fixing condition''.  Since these constraints remain second class after introducing the matter field, the number of gravitational degrees of freedom remains four in the phase space, i.e., two in the real space. This strategy has been successfully applied in  \cite{Aoki:2018zcv,Aoki:2018brq} by adding to the Hamiltonian a gauge fixing term ${\cal H}_{\rm{gf}}$ which is, by definition, not commuting with the Hamiltonian constraint. In our case, one need to introduce a gauge fixing term which does not commute
with the ${\cal H}_0$ or equivalently ${\cal H}_{gr}$. Following \cite{Aoki:2018brq}, one could think about adding to the total Hamiltonian a term 
which imposes, using a Lagrange multiplier, a new constraint either of the form $ \partial_i {\cal S}\approx 0$ or of the form ${\cal S} \approx 0$ where ${\cal S}$ is a three-dimensional scalar, such that, together with ${\cal H}_0$, they form a pair of second class constraints while the invariance under space-like diffeomorphisms is preserved. The coupling to matter (particularly the choice of ${\cal H}_{\rm{gf}}$) needs to be studied in great details and goes beyond the scope of the present work. For this reason, we leave this analysis for future investigations. }

\subsubsection{Cosmology}
To illustrate the difference between $f({\cal H})$ theories and general relativity, we consider simple examples. First, let us study the cosmology of these theories in the presence of
a perfect fluid (of density $\rho$ and pressure $p$) which
corresponds to taking a time dependent lapse function $N(t)$, a vanishing shift vector $N^i=0$, homogeneous and isotropic spatial metric and momenta as follows
\bea
h_{ij} \; = \; a^2(t) \delta_{ij} \, , \qquad \pi^{ij} \; = \; b(t) \delta^{ij} \, .
\eea
Here we assume that the spatial slices are flat.
{To make the dynamics in the cosmological sector more interesting, we consider the coupling to matter in the form a perfect fluid, as we have said previously. In that case, contrary to the generic situation, we do not really need
an explicit form of ${\cal H}_{\rm{gf}}$. Indeed, if the gauge-fixing condition is of the form $\partial_i {\cal S} \approx 0$, then 
it is trivially satisfied by FLRW space-time with the (space-independent) time
reparametrization symmetry unbroken (namely, the lapse function is arbitrary).
On the other hand, if the gauge condition of the form ${\cal S} \approx 0$, it may
imply a specific choice of the lapse function if $\cal S$ involves a fixed function of time, for instance. In this case,
the (space-independent) time reparametrization symmetry is broken.
In any cases, the gauge fixing term does not explicitly show up in the equations of motion, and we can consider a minimal coupling to matter (as described in the footnote 3)
where the lapse is either free or fixed to a specific value. Hence, the deformed Hamiltonian constraint simplifies drastically and becomes}
\bea
\label{ModGcosmo}
f({\cal H}_{gr})+ 16\pi G_N \, \rho \; = \; 0 \qquad \text{with} \quad {\cal H}_{gr}=-\frac{3}{2} \left( \frac{b}{a}\right)^2 \, .
\eea
Furthermore, the Hamilton equations of motion reduces to
\bea
\dot{a} = - \frac{N f'({\cal H}_{gr})}{2} b \, , \qquad \dot{b} = - \frac{N}{2} a \left[ f({\cal H}_{gr}) + f'({\cal H}_{gr}) \left(\frac{b}{a} \right)^2- 16\pi G_N \,  p \right]\, .
\eea
{Notice that FLRW cosmology could also be analyzed starting from the Lagrangian \eqref{minimalLag}
where $C$ has been defined by the relation \eqref{constraintC}. The result is, as expected, the same as in the
Hamiltonian formalism.} 

In general, the Friedmann equations are strongly modified compared to the classical ones. To write them,
it is useful to introduce 
\bea
F(\rho)=f^{-1}(-16\pi G_N \, \rho) \quad \Longrightarrow \quad f'({\cal H}_{gr}) = - \frac{16\pi G_N}{F'(\rho)} \, ,
\eea
in order to reformulate the previous three equations (with $N=1$) equivalently as follows
\bea
\left( \frac{b}{a}\right)^2 = -\frac{2}{3} F(\rho)  \, , \quad
b =  \frac{F'(\rho)}{8 \pi G_N} \dot{a} \, , \quad
\dot{b}= 8\pi G_N {a} \left[ \rho+ p - \frac{2 F(\rho)}{3 F'(\rho)} \right] \, ,
\eea
which lead to the following modified Friedmann equations
\bea
&& H^2=  -\frac{2}{3}(8\pi G_N)^2\frac{F(\rho)}{[F'(\rho)]^2} \, , \\
&& F'(\rho) \frac{\ddot a}{a} - 3H^2 F''(\rho)(\rho+p) = (8\pi G_N)^2 \left[ \rho+p - \frac{2F(\rho)}{3F'(\rho)}\right]  \, ,
\eea
whereas the conservation equation for the fluid remains unchanged
\bea
\dot{\rho} + 3H (\rho+p) \; = \; 0 \, .
\eea
When $f(x)=x$, one immediately recover the usual Friedmann equations.
Furthermore, in vacuum (when $\rho=0=p$), these equations admit a self-accelerating solution if $F(0) < 0$. 
This is for instance the case for
\bea
\label{examplefx}
f(x) \;  = \;  x \left(1 - \frac{x}{2\Lambda} \right) \qquad \Longrightarrow \quad
F(\rho) \; = \; {\Lambda} \left[ -1 \pm \sqrt{1- 32 \pi G_N  \rho/\Lambda}\right]
\eea
where $\Lambda$ is a non-negative constant. The function $F(\rho)$ has two branches, and the minus branch, which is such that
$F(0)=-2\Lambda <0$, admits a self-accelerating solution in vacuum with cosmological constant $\Lambda$. This result has a simple
interpretation. Indeed, in vacuum, the modified Hamiltonian constraint reduces to $f({\cal H}_{gr})=0$ whose solutions fall into two branches:
${\cal H}_{gr}=0$ which corresponds to general relativity with no cosmological constant and ${\cal H}_{gr}=2\Lambda$ which corresponds to general
relativity with a cosmological constant. In general, any deformation of general relativity associated to $f(x)$ admits a self accelerating solution  if $f(x)=0$
admits a non-negative solution $x_0$.

{Notice that in the absence of matter,
the FLRW background reduces to a de Sitter spacetime and that the
analysis of scalar perturbations about the de Sitter background
(without matter nor gauge fixing term) confirms that no scalar modes
are propagating in these theories.}

\section{Conclusion}
In this paper, we constructed theories of minimally modified  gravity (MMG) from a Hamiltonian point of view. 
To illustrate the construction, we started in section \ref{Maxwell}  with a complete study of minimally modified Mawxell theories which propagates 2 (vectorial) degrees of freedom in the 
4-dimensional Minkowski space-time. Maxwell theory provides us with a simpler but very interesting context to present the main ingredients that enter in the construction of minimally modified gravity theories from a Hamiltonian point of view. Then, we considered the most interesting case of gravity.
We started with the phase space of general relativity parametrized with 10 pairs of canonically conjugate variables (the metric components and their momenta) and whose dynamics is governed by the Hamiltonian and vectorial constraints. We modified the theory in such a way that, first, the lapse function function and the shift vector remain non-dynamical (i.e. with vanishing conjugate momenta), second, the theory is still invariant under 3D diffeomorphisms, and third the theory propagates only two tensorial degrees of freedom. We found that these three requirements lead to a Hamiltonian of the form \eqref{HMMG} with the condition \eqref{selfcommutation}. 

We showed that these MMG theories encompass the so-called 
cuscuton theories (in the unitary gauge) which are (higher derivative) scalar-tensor theories with only two tensorial modes. In these theories, the scalar degree of freedom is in fact
a shadow mode \cite{DeFelice:2018mkq} and thus  does not propagate. Notice that our construction naturally extends the cuscuton models to non-local theories which involve
infinite spatial derivatives.
We also found a particularly interesting and simple novel class of MMG whose Hamiltonian differs from the Hamiltonian of general relativity by the fact that the Hamiltonian constraint ${\cal H}_{gr}$ has been replaced by $f({\cal H}_{gr})$ where $f$ is an arbitrary function. We dubbed them $f({\cal H})$-theories . 

The class of $f({\cal H})$-theories opens numerous new windows in cosmology and in astrophysics.  We have quickly studied cosmological solutions 
for a generic choice of function $f(x)$, but it would be interesting to make a systematic analysis of cosmological perturbations and of
the constraints that observations put on these theories if they account for dark energy. 
For that, it is important to first understand in details how to consistently couple matter in these theories following the analysis of \cite{Aoki:2018zcv,Aoki:2018brq}. This would also allow us to study, for instance, the structure of stars in these theories and to see how Newton laws are modified in this framework.
 From a more formal point of view, we are curious to understand the relations and the differences 
with the very well-studied $f(R)$ or $f(R,T)$ theories. We hope to investigate all these questions in the future...

\subsection*{Acknowledgments}
We want to thank David Langlois for very interesting and useful discussions. 
The work of S.M. was supported by Japan Society for the Promotion of
Science (JSPS) Grants-in-Aid for Scientific Research (KAKENHI) No.
17H02890, No. 17H06359, and by World Premier International Research
Center Initiative (WPI), MEXT, Japan. He is grateful to Institut Denis
Poisson for hospitality during his stay.

\bibliographystyle{utphys}
\bibliography{Min1}

\end{document}